\documentclass[useAMS,usenatbib]{mn2e}
\usepackage[dvips]{color}
\usepackage{amssymb,graphicx,multirow,txfonts}

\newcommand{\etal}{et~al.}
\newcommand{\PVdblt}{{\rm P}\kern 0.1em{\sc v}~$\lambda\lambda 1117, 1128$}
\newcommand{\CaIIdblt}{{\rm Ca}\kern 0.1em{\sc ii}~$\lambda\lambda 3934, 3969$}
\newcommand{\AlIIIdblt}{{\rm Al}\kern 0.1em{\sc iv}~$\lambda\lambda 1855, 1863$}
\newcommand{\CIVdblt}{{\rm C}\kern 0.1em{\sc iv}~$\lambda\lambda 1548, 1550$}
\newcommand{\MgIIdblt}{{\rm Mg}\kern 0.1em{\sc ii}~$\lambda\lambda 2796, 2803$}
\newcommand{\NVdblt}{{\rm N}\kern 0.1em{\sc v}~$\lambda\lambda 1238, 1242$}  
\newcommand{\SVIdblt}{{\rm S}\kern 0.1em{\sc vi}~$\lambda\lambda 933, 944$} 
\newcommand{\OVIdblt}{{\rm O}\kern 0.1em{\sc vi}~$\lambda\lambda 1031, 1037$} 
\newcommand{\SiIIdblt}{{\rm Si}\kern 0.1em{\sc ii}~$\lambda\lambda 1190, 1193$} 
\newcommand{\SiIVdblt}{{\rm Si}\kern 0.1em{\sc iv}~$\lambda\lambda 1393, 1402$} 
\newcommand{\AlI}{\hbox{{\rm Al}\kern 0.1em{\sc i}}}
\newcommand{\AlII}{\hbox{{\rm Al}\kern 0.1em{\sc ii}}}
\newcommand{\AlIII}{{\hbox{\rm Al}\kern 0.1em{\sc iii}}}
\newcommand{\CaII}{\hbox{{\rm Ca}\kern 0.1em{\sc ii}}}
\newcommand{\CII}{\hbox{{\rm C}\kern 0.1em{\sc ii}}}
\newcommand{\CIIe}{\hbox{{\rm C$^{\ast}$}\kern 0.1em{\sc ii}}}
\newcommand{\CIII}{\hbox{{\rm C}\kern 0.1em{\sc iii}}}
\newcommand{\CIV}{\hbox{{\rm C}\kern 0.1em{\sc iv}}}
\newcommand{\CV}{\hbox{{\rm C}\kern 0.1em{\sc v}}}
\newcommand{\HI}{\hbox{{\rm H}\kern 0.1em{\sc i}}}
\newcommand{\HII}{\hbox{{\rm H}\kern 0.1em{\sc ii}}}
\newcommand{\Lya}{\hbox{{\rm Ly}\kern 0.1em$\alpha$}}
\newcommand{\Lyb}{\hbox{{\rm Ly}\kern 0.1em$\beta$}}
\newcommand{\Lyg}{\hbox{{\rm Ly}\kern 0.1em$\gamma$}}
\newcommand{\Lyd}{\hbox{{\rm Ly}\kern 0.1em$\delta$}}
\newcommand{\HeI}{\hbox{{\rm He}\kern 0.1em{\sc i}}}
\newcommand{\HeII}{\hbox{{\rm He}\kern 0.1em{\sc ii}}}
\newcommand{\FeI}{\hbox{{\rm Fe}\kern 0.1em{\sc i}}}
\newcommand{\FeII}{\hbox{{\rm Fe}\kern 0.1em{\sc ii}}}
\newcommand{\FeIII}{\hbox{{\rm Fe}\kern 0.1em{\sc iii}}}
\newcommand{\MnII}{\hbox{{\rm Mn}\kern 0.1em{\sc ii}}}
\newcommand{\MgI}{\hbox{{\rm Mg}\kern 0.1em{\sc i}}}
\newcommand{\MgII}{\hbox{{\rm Mg}\kern 0.1em{\sc ii}}}
\newcommand{\MgIII}{\hbox{{\rm Mg}\kern 0.1em{\sc iii}}}
\newcommand{\NI}{\hbox{{\rm N}\kern 0.1em{\sc i}}}
\newcommand{\NII}{\hbox{{\rm N}\kern 0.1em{\sc ii}}}
\newcommand{\NIII}{\hbox{{\rm N}\kern 0.1em{\sc iii}}}
\newcommand{\NV}{\hbox{{\rm N}\kern 0.1em{\sc v}}}
\newcommand{\OVI}{\hbox{{\rm O}\kern 0.1em{\sc vi}}}
\newcommand{\OI}{\hbox{{\rm O}\kern 0.1em{\sc i}}}
\newcommand{\OII}{\hbox{[{\rm O}\kern 0.1em{\sc ii}]}}
\newcommand{\OIII}{\hbox{[{\rm O}\kern 0.1em{\sc iii}]}}
\newcommand{\OIV}{\hbox{{\rm O}\kern 0.1em{\sc iv}]}}
\newcommand{\SI}{{\rm S}\kern 0.1em{\sc i}}
\newcommand{\SIV}{{\rm S}\kern 0.1em{\sc iv}}
\newcommand{\SVI}{{\rm S}\kern 0.1em{\sc vi}}
\newcommand{\SiI}{\hbox{{\rm Si}\kern 0.1em{\sc i}}}
\newcommand{\SiII}{\hbox{{\rm Si}\kern 0.1em{\sc ii}}}
\newcommand{\SiIII}{\hbox{{\rm Si}\kern 0.1em{\sc iii}}}
\newcommand{\SiIV}{\hbox{{\rm Si}\kern 0.1em{\sc iv}}}
\newcommand{\SII}{\hbox{{\rm S}\kern 0.1em{\sc ii}}}
\newcommand{\SIII}{\hbox{{\rm S}\kern 0.1em{\sc iii}}}
\newcommand{\NaI}{\hbox{{\rm Na}\kern 0.1em{\sc i}}}
\newcommand{\TiII}{\hbox{{\rm Ti}\kern 0.1em{\sc ii}}}
\newcommand{\ZnII}{\hbox{{\rm Zn}\kern 0.1em{\sc ii}}}
\newcommand{\CrII}{\hbox{{\rm Cr}\kern 0.1em{\sc ii}}}
\newcommand{\kms}{\hbox{km~s$^{-1}$}}

\newcommand{\NaID}{\hbox{{\rm Na}\kern 0.1em{\sc i}}\kern 0.05em{\rm D}}
\newcommand{\MgIb}{\hbox{{\rm Mg}\kern 0.1em{\sc i}}\kern 0.05em{\rm b}}
\def\aj{{AJ}}                   
             
\def\apj{{ApJ}}                 
\def\apjl{{ApJ}}                
\def\apjs{{ApJS}}

\def\aap{{A\&A}}

\def\mnras{{MNRAS}}

\def\pasp{{PASP}}

\title[The CGM of AGNs]{Probing the circumgalactic medium of
  active galactic nuclei with background quasars}

\author[G. G. Kacprzak et al.]{Glenn G. Kacprzak,$^{1,2}$\thanks{gkacprzak@astro.swin.edu.au} Christopher W. Churchill,$^{3}$ Michael T. Murphy,$^{1}$ Jeff Cooke$^{1}$\\
$^{1}$ Centre for Astrophysics and Supercomputing, Swinburne University of Technology, PO Box 218, Victoria 3122, Australia\\
$^{2}$ Australian Research Council Super Science Fellow\\
$^{3}$ Department of Astronomy, New Mexico State University, Las Cruces, NM 88003
}

\begin{document}
\date{}

\pagerange{\pageref{firstpage}--\pageref{lastpage}} \pubyear{2014}

\maketitle

\label{firstpage}

\begin{abstract}
  \noindent We performed a detailed study of the extended cool gas,
  traced by {\MgII} absorption [$W_r(2796)\geq0.3$~{\AA}], surrounding
  14 narrow-line active galactic nuclei (AGNs) at $0.12\leq z\leq
  0.22$ using background quasar sight-lines. The background quasars
  probe the AGNs at projected distances of $60\leq D\leq265$~kpc. We
  find that, between $100\leq D\leq200$~kpc, AGNs appear to have
  lower {\MgII} gas covering fractions (0.09$^{+0.18}_{-0.08}$) than
  quasars (0.47$^{+0.16}_{-0.15}$) and possibly lower than in active
  field galaxies (0.25$^{+0.11}_{-0.09}$). We do not find a
  statistically significant azimuthal angle dependence for the {\MgII}
  covering fraction around AGNs, though the data hint at one. We also
  study the `down-the-barrel' outflow properties of the AGNs
  themselves and detect intrinsic {\NaID} absorption in 8/8 systems
  and intrinsic {\MgII} absorption in 2/2 systems, demonstrating that
  the AGNs have significant reservoirs of cool gas.  We find that 6/8
  {\NaID} and 2/2 {\MgII} intrinsic systems contain blueshifted
  absorption with $\Delta v>50$~{\kms}, indicating outflowing gas. The
  2/2 intrinsic {\MgII} systems have outflow velocities a factor of
  $\sim4$ higher than the {\NaID} outflow velocities.  Our results are
  consistent with AGN-driven outflows destroying the cool gas within
  their halos, which dramatically decreases their cool gas covering
  fraction, while star-burst driven winds are expelling cool gas into
  their circumgalactic media (CGM).  This picture appears contrary to
  quasar--quasar pair studies which show that the quasar CGM contains
  significant amounts of cool gas whereas intrinsic gas found
  `down-the-barrel' of quasars reveals no cool gas. We discuss how
  these results are complementary and provide support for the AGN
  unified model.
\end{abstract}


\section{Introduction}

Galactic outflows originating in regions of high star formation
surface-density and in active galactic nuclei (AGNs) likely play a
significant role in regulating the metal content of galaxies and are
probably fully responsible for the chemical enrichment of the
intergalactic medium (IGM)
\citep[e.g.][]{oppenheimer10}. Understanding the distribution and
extent of the cool gas surrounding galaxies can aid in constraining
the metal contribution via winds to the CGM and IGM.  Though there are
many studies of outflows originating from star-forming galaxies
\citep{tremonti07,zibetti07,martin09,weiner09,noterdaeme10,
  rubin10,steidel10,bordoloi11,coil11,kacprzak11c,nestor11,menard12,martin12,rubin13,bordoloi13},
we are only beginning to understand the effect and distribution of gas
surrounding AGNs
\citep[e.g.][]{heckman00,martin05,rupke05a,rupke05b}.

Recent work has shown that large gas reservoirs surround quasars with
a covering fraction of 60--80\% and a physical extent of 200~kpc
\citep{hennawi06,bowen06,prochaska09,tytler09,farina13,prochaska13,farina14},
similar to what is observed for inactive field galaxies
\citep{kacprzak08,chen10,nielsen13,churchill13a,churchill13b}. The
cool gas surrounding field galaxies, as traced by {\MgII} absorption,
exhibits an anisotropic distribution whereby most gas is located along
their projected major and minor axes, which is interpreted to be due
to accretion and outflows, respectively
\citep{kacprzak11b,bordoloi11,bouche12,bordoloi12,kacprzak12}.
Quasars also exhibit an anisotropic gas distribution possibly caused
by the intense AGN-driven ionizing radiation that heats the gas to
temperatures of $\sim 10^5$K and destroys cool gas clouds out to a few
hundred kiloparsecs \citep{chelouche08}. \citet{prochaska13} analyzed
74 quasar--quasar pairs with projected separations of $D< 300$~kpc and
showed that, although quasars exhibit a high covering fraction of cool
gas as traced by {\HI} and {\CII}, cool gas is rarely detected along
the jet (`down-the-barrel') of the quasar itself. They suggest that
the background quasar sight-lines intercept gas that is shadowed from
the ionizing radiation of the quasar jet. Similar anistropy has been
observed in {\HI} \citep{hennawi07,prochaska09} and {\MgII}
\citep{bowen06,farina13,farina14}. The distribution of the CGM of
AGNs, like Seyfert galaxies, has yet to be explored.

If the AGN radiation field is responsible for destroying cool gas,
then the orientation of the AGN-driven outflows with respect to the
background quasar sight-line may dictate the presence, strength and
frequency of the absorption. The inclination of the AGN-driven
outflows are primarily constrained by the unified AGN model
\citep[see][ and references therein]{bianchi2012}. The type of AGN
observed may be dependent on the observers' vantage point, or
essentially the AGN dust torus inclination. Quasars and blazers are
typically viewed down-the-barrel of the AGN jet (face-on torus), while
broad-line Seyfert~I objects are viewed at a slightly higher
inclination and narrow-line Seyfert~IIs are viewed with an edge-on
torus. If one were to probe quasars or blazars with background
quasars, the dusty torus is in the plane of the sky (face-on and
orthogonal to the jet) and provides the maximum cross-section of
radiation-shielded extended halo gas, yielding the highest covering
fraction of cool halo gas. Therefore, we would expect that the cool
gas covering fraction surrounding AGNs should decrease as the
inclination of the dusty torus decreases from face-on to edge-on for
quasars, Seyfert~Is and Seyfert~IIs, respectively.

Motivated by the idea that the gas covering fraction of AGNs may
change with viewing angle (according to the unified model), we target
a sample of 14 narrow-line AGN (Seyfert IIs) to examine their cool CGM
gas covering fractions. We target $z\sim 0.15$ AGN so that we are able
to detect the {\MgIIdblt} and {\NaID} $\lambda\lambda5892, 5897$
absorption doublets from ground-based facilities and we are able to
resolve the morphological properties of their host galaxies in
ground-based images. We investigate the cool gas covering fractions
and azimuthal angle dependencies. We show that AGN exhibit intrinsic
absorption, unlike quasars, and we quantify their outflow properties.
In Section~\ref{sec:data} we present our sample and data reduction. In
Section~\ref{sec:results}, we present the gas covering fractions and
azimuthal dependencies. present clear signs of intrinsic
(down-the-barrel) cool gas outflows originating from ongoing
star-formation rather than the jets and quantify their outflow
properties.  In Section~\ref{sec:dis}, we discuss what can be inferred
from the results and how it arises naturally from the AGN unified
model.  Concluding remarks are offered in
Section~\ref{sec:conclusion}. Throughout we adopt an H$_{\rm
  0}=70$~\kms Mpc$^{-1}$, $\Omega_{\rm M}=0.3$, $\Omega_{\Lambda}=0.7$
cosmology.

%%%%%%%%%%%%%%%%%%%%%%%%%%%%%%%%%%%%%%%%%%%%%%%%%%%%%%%%%%%%%%%%%%%%%%%%%%%%%%
\begin{figure*}
\includegraphics[angle=0,scale=0.63]{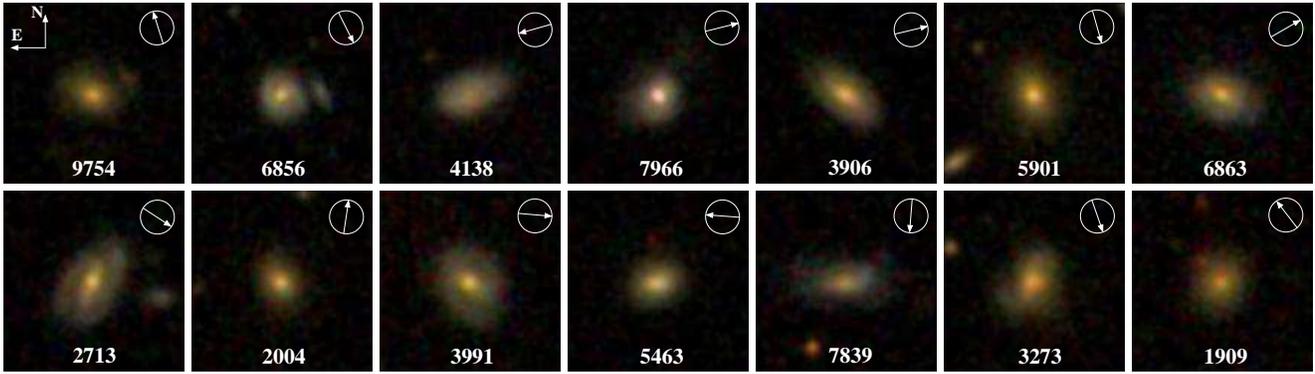}
\caption{$gri$-band SDSS color images of the 14 foreground narrow-line
  AGN (Seyfert IIs). The direction of the background quasar sight-line
  is indicated by the arrow in the upper-right corner of the
  images. The AGN--quasar separations range between 63--265~kpc. The
  AGN--quasar pair IDs are located on the bottom of each image. Note
  that the galaxies tend to have low inclination angles, with average
  inclination of $i=51\pm12$ degrees. }
\label{fig:stamps}
\end{figure*}
%%%%%%%%%%%%%%%%%%%%%%%%%%%%%%%%%%%%%%%%%%%%%%%%%%%%%%%%%%%%%%%%%%%%%%%%%%%%%%

\begin{table*}
\begin{center}
  \caption{Keck--I/LRIS foreground narrow-line AGN and background
    quasar observations. The table columns are (1) the quasar name,
    (2) the quasar RA, (3) the quasar DEC, (4) the AGN and quasar pair
    name, (5) the quasar redshift, (6) the quasar SDSS $g$-band
    apparent magnitude, (7) the observation date, and (8) the total
    integration time in seconds.} \vspace{-0.5em}
\label{tab:qsospec}
{\footnotesize\begin{tabular}{lcccrccr}\hline
SDSS                        &  RA$_{QSO}$& DEC$_{QSO}$  & Pair &   $z_{QSO}$ & $m_g$ & Date (UT)    & Exposure\\  
Quasar Name                 &  (J2000)     &(J2000)        & Name &               &       &              &   (sec.)\\\hline
SDSS J230632.06$+$004611.7  & 23:06:32.06 & $+$00:46:11.78 & 9754 &    0.57072   &   19.8 & Sep. 18 2009 & 1880  \\
SDSS J133216.60$+$020634.3  & 13:32:16.60 & $+$02:06:34.39 & 6856 &    0.52960   &   19.2 & Feb. 12 2010 & 3720   \\
SDSS J101133.17$+$011451.1  & 10:11:33.17 & $+$01:14:51.17 & 4138 &    0.37035   &   19.9 & Feb. 13 2010 & 4960   \\
SDSS J144732.28$+$610722.1  & 14:47:32.28 & $+$61:07:22.19 & 7966 &    0.36189   &   19.6 & Feb. 13 2010 & 4300   \\
SDSS J095105.22$+$000049.2  & 09:51:05.22 & $+$00:00:49.29 & 3906 &    0.87571   &   19.3 & Feb. 13 2010 & 4960   \\
SDSS J123844.41$-$001140.2  & 12:38:44.41 & $-$00:11:40.22 & 5901 &    0.87540   &   18.8 & Feb. 13 2010 & 3720   \\
SDSS J133248.52$+$014250.3  & 13:32:48.52 & $+$01:42:50.35 & 6863 &    0.24717   &   19.3 & Feb. 12 2010 & 5120   \\
SDSS J080626.99$+$465857.1  & 08:06:26.99 & $+$46:58:57.12 & 2713 &    1.35435   &   19.7 & Feb. 13 2010 & 4960  \\
SDSS J032605.40$-$073242.2  & 03:26:05.40 & $-$07:32:42.23 & 2004 &    1.05407   &   19.6 & Feb. 12 2010 & 3720   \\
SDSS J095701.58$+$023857.3  & 09:57:01.58 & $+$02:38:57.32 & 3991 &    1.07677   &   19.3 & Feb. 12 2010 & 3720  \\
SDSS J120210.73$+$005520.4  & 12:02:10.73 & $+$00:55:20.44 & 5463 &    1.18228   &   19.4 & Feb. 12 2010 & 3720  \\
SDSS J143729.16$+$024404.2  & 14:37:29.16 & $+$02:44:04.21 & 7839 &    1.30369   &   19.8 & Feb. 13 2010 & 3180  \\
SDSS J085755.38$+$531145.3  & 08:57:55.38 & $+$53:11:45.32 & 3273 &    1.57320   &   19.6 & Feb. 12 2010 & 3720   \\
SDSS J031531.50$-$074002.7  & 03:15:31.50 & $-$07:40:02.71 & 1909 &    1.05373   &   20.0 & Feb. 13 2010 & 2480   \\\hline 
\end{tabular}}
\end{center}
\end{table*}

\begin{table*}
\begin{center}
  \caption{The foreground AGN and absorption properties. The table
    columns are (1) the AGN--quasar pair name, (2) the SDSS AGN
    redshift, (3) the AGN RA, (4) the AGN DEC, (5) the AGN SDSS
    $g$-band magnitude, (6) the inclination of the host-AGN, (7) The
    orientations of the quasar sight-lines with respect to the
    projected major (x-axis, $\Phi=0$~degrees) and minor (y-axis,
    $\Phi=90$~degrees) axis of the AGNs, (7) the projected separation
    between background quasar and foreground AGN pair, (8) the {\MgII}
    $\lambda 2796$ rest-frame equivalent width with 1$\sigma$ errors
    or 3$\sigma$ limits, (8) the {\MgII} $\lambda 2803$ rest-frame
    equivalent width with 1$\sigma$ errors or 3$\sigma$ limits, (9)
    the {\NaID} rest-frame equivalent width limits (3$\sigma$), and
    (10) the {\MgII} $\lambda 2796$ absorption-line redshift.}
  \vspace{-0.5em}
\label{tab:qsoresults}
{\footnotesize\begin{tabular}{lccccllrcccc}\hline
Pair &$z_{AGN}$  &  RA$_{AGN}$   & DEC$_{AGN}$   &$m_g$ &$\phantom{000}$$i$ &$\phantom{000}$$\Phi$& $D$$\phantom{0}$ & $W_r(2796)$   &   $W_r(2803)$&     $W_r($NaID$)$ &   $z_{abs}$\\
Name &          &   (J2000)     & (J2000)      &      & (degrees)            &    (degrees)       & (kpc)   &  (\AA)         & (\AA)        &   (\AA)      &             \\\hline 
9754 & 0.193216 &  23:06:31.674 &$+$00:45:53.0 & 18.7 &$54.5_{-7.2}^{+5.8}$   &   $48.9_{-9.8}^{+13.9}$&  62.6  &  1.05$\pm$0.10 &0.86$\pm$0.10 &     $\cdots$  &   0.193263\\
6856 & 0.129467 &  13:32:18.116 &$+$02:07:19.5 & 18.0 &$34.6_{-10.5}^{+24.2}$ &  $68.5_{-22.7}^{+20.5}$& 117.6  &    $<$0.27     &     $<$0.27  &     $<$0.08  &            \\
4138 & 0.121781 &  10:11:29.309 &$+$01:15:07.7 & 18.1 &$59.5_{-2.5}^{+3.9}$   &  $83.5_{-3.9}^{+3.7}$  & 134.1  &    $<$0.23     &     $<$0.23  &     $<$0.10  &            \\
7966 & 0.136349 &  14:47:39.330 &$+$61:06:56.1 & 17.5 &$27.3_{-8.3}^{+14.3}$  &  $38.0_{-38.7}^{+31.5}$& 140.2  &    $<$0.30     &     $<$0.32  &    $\cdots$  &            \\
3906 & 0.132690 &  09:51:09.084 &$+$00:00:35.4 & 18.0 &$62.5_{-2.6}^{+3.0}$   &  $36.1_{-3.7}^{+4.7}$  & 141.8  &    $<$0.11     &     $<$0.11  &     $<$0.09  &            \\
5901 & 0.139482 &  12:38:45.634 &$-$00:10:36.4 & 18.3 &$52.0_{-6.3}^{+6.3}$   &  $87.4_{-10.9}^{+8.3}$ & 164.8  &    $<$0.13     &     $<$0.13  &     $<$0.13  &            \\
6863 & 0.165819 &  13:32:51.879 &$+$01:42:21.2 & 17.9 &$62.6_{-23.7}^{+13.6}$ &  $35.8_{-27.2}^{+16.5}$& 164.8  &    $<$0.20     &     $<$0.20  &   $\cdots$  &            \\
2713 & 0.124705 &  08:06:32.257 &$+$46:59:48.6 & 17.8 &$58.9_{-2.6}^{+3.4}$   &  $\phantom{0}8.8_{-3.5}^{+3.5}$   & 168.2  & 1.37$\pm$0.10  &1.60$\pm$0.09 &    $\cdots$   &  0.124438 \\   
2004 & 0.156050 &  03:26:06.049 &$-$07:33:46.7 & 18.6 &$48.2_{-4.9}^{+4.3}$   &  $59.2_{-6.8}^{+8.6}$  & 176.4  &    $<$0.19     &     $<$0.19  &    $<$0.08   &            \\
3991 & 0.127130 &  09:57:06.861 &$+$02:39:04.0 & 17.9 &$39.5_{-4.1}^{+5.3}$   &  $36.2_{-9.7}^{+10.8}$ & 182.2  &    $<$0.17     &     $<$0.17  &    $\cdots$  &            \\
5463 & 0.163679 &  12:02:06.463 &$+$00:55:15.6 & 18.3 &$43.4_{-7.1}^{+9.8}$   &  $68.6_{-11.6}^{+10.5}$& 182.6  &    $<$0.17     &     $<$0.17  &     $<$0.07  &            \\
7839 & 0.179098 &  14:37:28.817 &$+$02:45:08.7 & 18.3 &$67.0_{-3.3}^{+3.5}$   &  $\phantom{0}5.8_{-11.3}^{+8.4}$  & 198.4  &    $<$0.21     &     $<$0.21  &     $<$0.11  &            \\
3273 & 0.163778 &  08:57:57.341 &$+$53:13:09.5 & 18.2 &$49.9_{-7.9}^{+5.1}$   &  $48.9_{-7.1}^{+7.8}$  & 244.1  &    $<$0.28     &     $<$0.28  &    $\cdots$  &            \\
1909 & 0.198855 &  03:15:28.231 &$-$07:41:06.2 & 18.8 &$32.3_{-9.6}^{+9.5}$   &  $41.7_{-20.9}^{+16.9}$& 265.2  &    $<$0.18     &     $<$0.18  &    $<$0.11   &            \\\hline
\end{tabular}}
\end{center}
\end{table*}

\section{GALAXY SAMPLE AND DATA ANALYSIS}
\label{sec:data}

\subsection{Sample Selection}

To probe the gas environment around AGN we have searched for AGN that
lie in the foreground of background quasars.  We cross-correlated a
catalogue of $\sim$12,000 AGN that are spectroscopically classified
using SDSS (DR2) spectra \citep{hao05} with spectroscopically
classified quasars in SDSS (DR7). Out of the 77 foreground
AGN--background quasar pairs identified, we applied the following
selection criteria: (1) The projected separations are less than
400~kpc; (2) A redshift separation of $\Delta v>20,000$~{\kms} to
avoid confusion between the foreground AGN {\MgII} absorption and the
intrinsic {\MgII} absorption from the quasar; (3) Both the foreground
AGN and the background quasar have a SDSS $g$-band apparent magnitude
of $m_g<20$; (4) The {\MgII} absorption occurs redward of the quasar
{\Lya} emission in order to avoid the {\Lya} forest.  A total of 31
foreground AGN--background quasar pairs meet these selection criteria.
We have observed 14 AGN--quasar pairs that were at optimal airmass
during our observing runs. The 14 foreground AGN presented here have a
redshift range determined by SDSS of $0.12\leq z \leq 0.22$. Typical
redshift errors are $\Delta z=0.00005$ ($\sim 15$~{\kms}). The
background quasars probe the foreground AGNs over an impact parameter
range of $60\leq D \leq 265$~kpc. All 14 objects are narrow-line AGN
(Seyfert~II).

\subsection{AGN \& Background Quasar Spectroscopy}\label{sec:qso_spec}

The combined wavelength coverage and ultra-blue sensitivity of
Keck/LRIS \citep{oke95,steidel04} is ideal for targeting {\MgII}
absorption \citep{barton09,kacprzak11a} and {\NaID} absorption
\citep{rupke05c} at $z=0.1-0.2$.  The Keck/LRIS foreground AGN and
background quasar spectra were obtained over three nights; one was
obtained in 2009 September and 13 were obtained in two nights in 2010
February. The LRIS slit was oriented such that the background quasar
and AGN fell within the slit. Details of the observations are
presented in Table~\ref{tab:qsospec}. We used the LRIS-B/Keck 1200
lines/mm grism, blazed at 3400~{\AA}, which covers a wavelength range
of 2910$-$3890\AA.  We used a 1.0$''$ slit that yields a dispersion of
0.24~{\AA} per pixel and provides a resolution of FWHM$\sim$1.6~{\AA}
($\sim$150~\kms). Integration times of 1880--5120 seconds were used,
depending on the magnitude of the quasar and the foreground AGN
redshift, providing 3$\sigma$ detection limits of $W_r(2796) \sim
0.3$\AA. We concurrently used the LRIS-R/Keck 1200 lines/mm grating
blazed at 7500~{\AA}, set to a central wavelength of 6314~\AA, which
covers a wavelength range of 5495$-$7133\AA.  The 1.0$''$ slit used
yields a dispersion of 0.40~{\AA} per pixel and provides a resolution
of FWHM$\sim$1.9~{\AA} ($\sim$85~\kms). The spectra were reduced using
the standard IRAF packages\footnote{IRAF is written and supported by
  the IRAF programming group at the National Optical Astronomy
  Observatories (NOAO) in Tucson, Arizona. NOAO is operated by the
  Association of Universities for Research in Astronomy (AURA), Inc.\
  under cooperative agreement with the National Science Foundation.}
and were corrected to the vacuum and heliocentric frame.

The quasar spectra were searched for {\MgII} and {\NaID} doublet
candidates using a detection significance level of 3~$\sigma$ for each
doublet member. Detection and significance levels follow the formalism
of \citet{schneider93} and \citet{archiveI}.  In addition, intervening
and intrinsic {\MgII} and {\CIV} absorption systems serendipitously
identified in the spectra that were not targeted in this survey are
listed in Table~\ref{tab:extra}. 

Analysis of the absorption profiles was performed using our own
graphics-based interactive software that uses the flux values in
individual pixels to measure the equivalent widths and the redshift of
the {\MgII} $\lambda 2796$ transition \citep{cv01}.  Flux weighted
absorption velocity widths were measured between the pixels where the
equivalent width per resolution element recovers to the $1~\sigma$
detection threshold \citep{weakI,archiveI}.  The redshift for each
{\MgII} and/or {\NaID} system is computed from the optical depth
weighted mean of the absorption profile. The statistical uncertainties
in the redshifts range between 0.00001--0.00009 ($\sim 3$--$30$~{\kms}
co-moving).

\subsection{AGN Images \& Models}

In Figure~\ref{fig:stamps} we show $gri$-band SDSS color images of
the 14 foreground AGN. The direction of the background quasar
sight-line is indicated by the arrow in the upper-right corner of the
images.

The AGN-host galaxy morphological parameters were determined by
applying the two-dimensional decomposition fitting program GIM2D
\citep{simard02} to the $r$-band images. The image
point-spread-function required by GIM2D was derived from nearby stars
in each image that were modeled using DAOPHOT
\citep{stetson87,stetson99}.

\section{Results}
\label{sec:results}

\subsection{Transverse Absorption -- Covering Fraction}

\begin{figure}
\includegraphics[angle=0,scale=0.43]{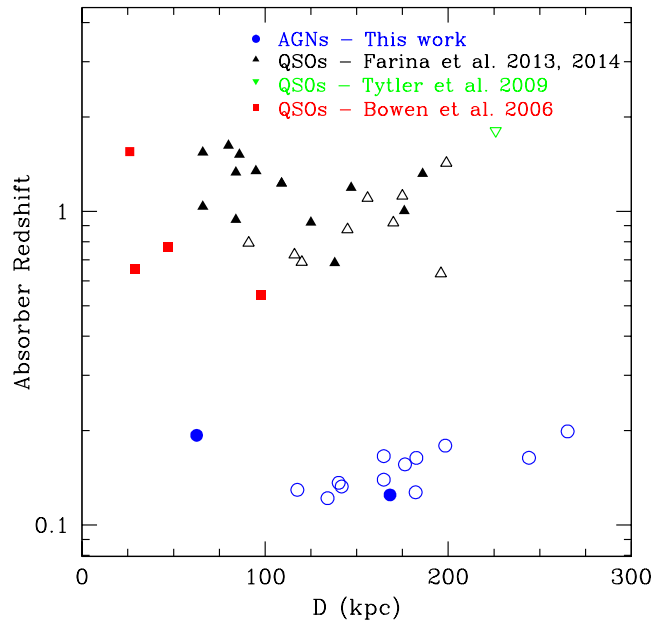}
\caption{Known {\MgII} absorption from quasar--quasar pairs
  (represented by squares and triangles) and foreground AGN--quasar
  pairs (represented by circles). The filled symbols are detected
  absorbers and the open symbols are non-detections with 3~$\sigma$
  equivalent width limits of $<0.3$~{\AA} for all four samples.}
\label{fig:samples}
\end{figure}

We present the first study of an AGN-selected sample shown in
Figure~\ref{fig:samples} and in Table~\ref{tab:qsoresults}. Our sample
of 14 background quasar and foreground AGN pairs probe an impact
parameter range of $60\leq D \leq 265$~kpc with a detection threshold
is $0.3$~{\AA} (3~$\sigma$). For our sample, we find two AGN with
absorption detected at $D=63$~kpc, with $W_r(2796)=1.1$~{\AA}, and at
$D=168$~kpc, with $W_r(2796)=1.4$~{\AA}.

In Figure~\ref{fig:samples} we present ``transverse'' {\MgIIdblt}
absorption doublet redshift versus impact parameter for our
AGN-selected sample and compare it with quasar-selected samples from
other published works.  The data presented in Figure~\ref{fig:samples}
have a detection limit of 0.3~{\AA} at the 3~$\sigma$
level. \citet{tytler09} studied 170 quasar--quasar pairs probing
distances out to 2~Mpc. Although this sample is large, the majority of
quasar--quasar pairs have velocity separations of
$\lesssim$20,000~{\kms}; it is therefore unclear if the absorption
detected is produced by gas surrounding the foreground quasar or by
out-flowing gas from the background quasar itself.  Only one quasar
pair resides within an impact parameter of 300 kpc and is separated by
$>20,000$~{\kms}. \citet{bowen06} and
\citet{farina13,farina14}\footnote{We have recomputed 3~$\sigma$
  detection limits for the quoted 2~$\sigma$ limits of
  \citet{farina13,farina14} and removed systems where revised
  3~$\sigma$ limits exceed 0.3~{\AA}.}  probe 4 and 22 $z\sim 1$
foreground quasars, respectively, for which $D<200$~kpc and the
velocity separation with respect to the background quasars is greater
than $20.000$~{\kms}.

\begin{figure}
\includegraphics[angle=0,scale=0.43]{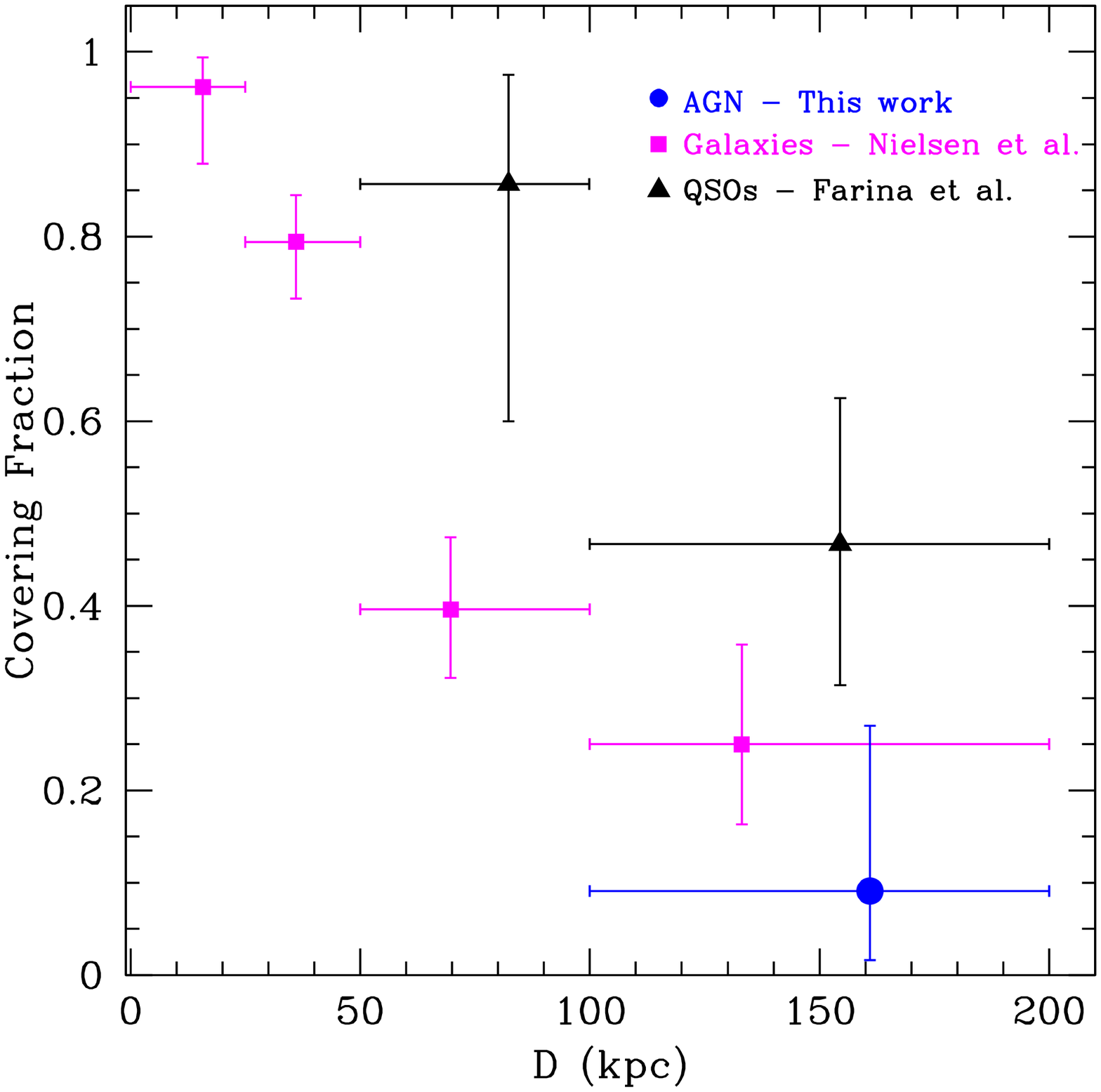}
\caption{The radial covering fraction profiles for AGN (this work),
  inactive field galaxies \citep{nielsen13} and quasars
  \citep{farina13,farina14}. All datasets have 3~$\sigma$ equivalent
  width limits of $>0.3$~{\AA}. The data points are located at the
  mean of the impact parameter distribution for each bin. AGNs have
  covering fractions that are consistent with field galaxies for
  $100\leq D \leq 200$~kpc. We have excluded our one datapoint at
  $D<100$~kpc from this plot since the errors are significantly larger
  and is consistent with both galaxies and quasars.}
\label{fig:cf}
\end{figure}

In Figure~\ref{fig:cf} we present the covering fraction profiles for
AGNs (this work), galaxies \citep{nielsen13} and quasars
\citep{farina13,farina14}.  The covering fractions for each sample are
determined per impact parameter bin, matching the bins of
\citet{nielsen13}. The plotted value indicates the mean $D$ per bin.
The covering fraction errors for each AGN, galaxy, quasar datasets are
derived from binomial statistics \citep{gehrels86}.

As found by \citet{farina14}, the covering fractions of quasars are
higher than those of isolated galaxies, which is likely due to the
richness and extent of the CGM around quasar-host galaxies. This
higher {\MgII} covering fraction is consistent with the high {\CII}
covering fraction ($\sim 50$\%) found for quasars \citep{prochaska13}.
However, AGN may exhibit similar or possibly lower covering fractions
as compared to field galaxies for $100\leq D \leq 200$~kpc.

Notwithstanding the obviously poor statistics at $D<100$~kpc, with one
data point (not included on the plot), the data are likely consistent
with previous works that AGN-type objects and galaxies have a high
covering fraction for $D<100$~kpc.

For 7 of the 14 quasar--AGN pairs, the spectra cover the {\NaID}
absorption doublet. However, no {\NaID} was detected down to a
$3\sigma$ equivalent width threshold of 0.13~{\AA}, which provides
further evidence for no cold gas surrounding the AGNs.

%%%%%%%%%%%%%%%%%%%%%%%%%%%%%%%%%%%%%%%%%%%%%%%%%%%%%%%%%%%%%%%%%%%%%%%%%%%%%%

% \citet{tytler09} has 10 quasar--quasar pairs with
%  $D<$300 kpc; however, only one has a velocity separation of $\Delta
%  v>20,000$~\kms.
%%%%%%%%%%%%%%%%%%%%%%%%%%%%%%%%%%%%%%%%%%%%%%%%%%%%%%%%%%%%%%%%%%%%%%%%%%%%%%
%%%%%%%%%%%%%%%%%%%%%%%%%%%%%%%%%%%%%%%%%%%%%%%%%%%%%%%%%%%%%%%%%%%%%%%%%%%%%%
%%%%%%%%%%%%%%%%%%%%%%%%%%%%%%%%%%%%%%%%%%%%%%%%%%%%%%%%%%%%%%%%%%%%%%%%%%%%%%

\subsection{Transverse Absorption -- AGN orientations}

\begin{figure}
\includegraphics[angle=0,scale=0.275]{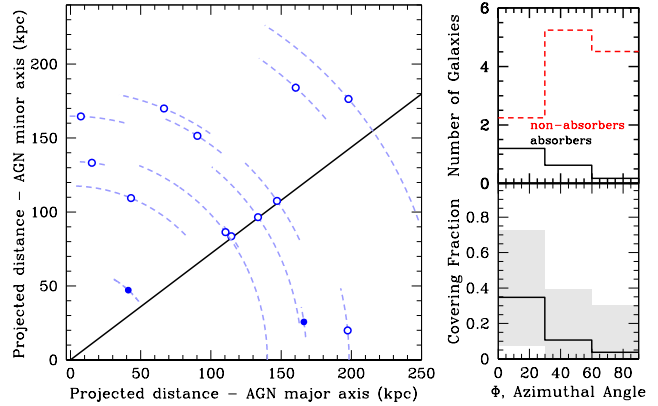}
\caption{(left) The orientations of the quasar sight-lines with
  respect to the projected major (x-axis, $\Phi=0$~degrees) and minor
  (y-axis, $\Phi=90$~degrees) axis of the AGNs. The filled symbols are
  detected {\MgII} absorbers and the open symbols are non-detections
  with 3~$\sigma$ equivalent width limits of $<0.3$~\AA. The
  1~$\sigma$ error-bars are determined from GIM2D model errors in
  determining the galaxy major axis position angle with respect to the
  quasar sight-line. The solid line represents a 50~degree
  half-opening angle typically found for AGN outflows. (right, top)
  The binned azimuthal angle mean probability distribution function
  for 2 absorbing (solid line) and 12 non-absorbing AGN (dashed line)
  produced using the technique described in \citet{kacprzak12}. The
  area of the histograms is normalized to the total number of galaxies
  in each sub-sample. (right, bottom) The azimuthal dependence of the
  covering fraction with shaded regions providing 1~$\sigma$ binomial
  errors. }
\label{fig:ori}
\end{figure}

In Figure~\ref{fig:ori}, we show the relative position of the quasar
sight-lines with respect to the host AGN projected major
($\Phi=0$~degrees) and minor ($\Phi=90$~degrees) axes.  The solid line
represents a typical half opening-angle determined for galactic-scale
winds and AGN outflows of 50 degrees
\citep{hjelm96,veilleux01,walter02,muller-Sanchez11,bordoloi12,kacprzak12,martin12}.
Note that most sight-lines probe within the expected region for
outflowing gas.

We attempted to determine if there is an azimuthal dependence for the
{\MgII} gas covering fraction shown in Figure~\ref{fig:ori}. Since the
azimuthal angle for each quasar varies in accuracy and can subtend
into several azimuthal angle bins, we apply the method of
\citet{kacprzak12} whereby we represent the measured azimuthal angles
and their uncertainties as uni-variate asymmetric Gaussians, which
creates an azimuthal angle probability distribution function (PDF) for
each galaxy. Combining the PDFs for all absorbers and non-absorbers
produces a mean PDF as a function of the azimuthal angle.

In Figure~\ref{fig:ori}, we present the binned mean azimuthal angle
($\Phi$) PDF for the 2 absorbing and 12 non-absorbing galaxies and
their covering fractions as a function of the azimuthal angle.  The
combined PDFs have been area-normalized to the total number of
galaxies in each sub-sample. The covering fraction is computed for
each azimuthal bin. The errors in the covering fractions are computed
using binomial statistics \citep{gehrels86}. The data are suggestive
that there could be an excess of absorbers along the galaxy projected
major axis ($\Phi=0$~degrees) compared to the minor axis
($\Phi=90$~degrees).  However, the current sample size is insufficient
to determine statistically if such a geometric dependence exists given
the binomial error distributions.

%%%%%%%%%%%%%%%%%%%%%%%%%%%%%%%%%%%%%%%%%%%%%%%%%%%%%%%%%%%%%%%%%%%%%%%%%%%%%%
%%%%%%%%%%%%%%%%%%%%%%%%%%%%%%%%%%%%%%%%%%%%%%%%%%%%%%%%%%%%%%%%%%%%%%%%%%%%%%

\subsection{Transverse Absorption -- Gas Kinematics}

\begin{figure}
\includegraphics[angle=0,scale=0.45]{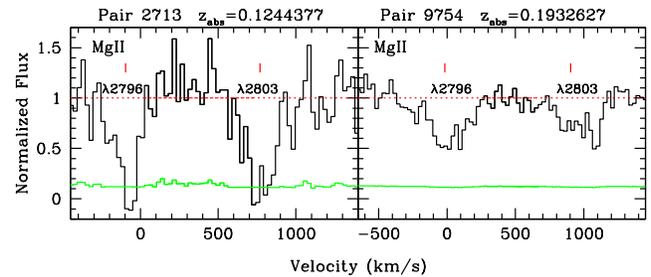}
\caption{Two {\MgIIdblt} absorption profiles associated with
  foreground AGNs at impact parameters of $D=168$~kpc and $D=63$~kpc
  from left to right, respectively. The velocity zero points are fixed
  to the redshifts of the AGN. The horizontal line (red) at unity is
  the continuum fit while the solid lines (green) are the 1~$\sigma$
  error spectra. }
\label{fig:profiles}
\end{figure}

In Figure~\ref{fig:profiles}, we present the two detected absorption
systems. The normalized flux of the {\MgIIdblt} doublet is plotted
with respect to the AGN velocity zero-point. The strongest absorber
(ID 2713) is the system at $D=168$~kpc and $z_{abs}=0.124438$. This
absorber has a {\MgII} rest-frame equivalent width of
$W_r(2796)=1.4$~{\AA} and a velocity offset of $-71$~{\kms} blueward
of the AGN redshift. The quasar sight-line probes the AGN along the
projected major axis.  The absorber has a full velocity width spanning
$272$~{\kms}.

The absorber at $D=63$~kpc and $z_{abs}=0.193263$ (ID 9754) has a
$W_r(2796)=$1.1~{\AA} and a velocity offset of $12$~{\kms} redward of
the AGN redshift with a full velocity width spanning
$355$~{\kms}. This absorber is located near the projected minor axis
of the galaxy.

The transverse absorption velocity offsets for these two systems,
relative to the AGN systemic velocities, are small and not suggestive
of outflows \citep[e.g.,][]{kacprzak10a}.

%%%%%%%%%%%%%%%%%%%%%%%%%%%%%%%%%%%%%%%%%%%%%%%%%%%%%%%%%%%%%%%%%%%%%%%%%%%%%%
%%%%%%%%%%%%%%%%%%%%%%%%%%%%%%%%%%%%%%%%%%%%%%%%%%%%%%%%%%%%%%%%%%%%%%%%%%%%%%

\subsection{Line-of-sight outflows}

The ionization potential of {\NaID} (5.14~eV) results in a peak in
absorption strength in the atmospheres of cool K--M stars
\citep[see][]{jacoby84}. In addition, it is further absorbed by cool
gas within, and surrounding, galaxies. For example, {\NaID} has been
detected as entrained gas within galactic-scale outflows with
velocities ranging between 100--1000~{\kms}
\citep[e.g.][]{heckman00,martin05,rupke05a,rupke05b}.  These outflows
are measured by observing velocity offsets between galaxy nebular
emission lines and ISM absorption lines.

\begin{figure}
\includegraphics[angle=0,scale=1.15]{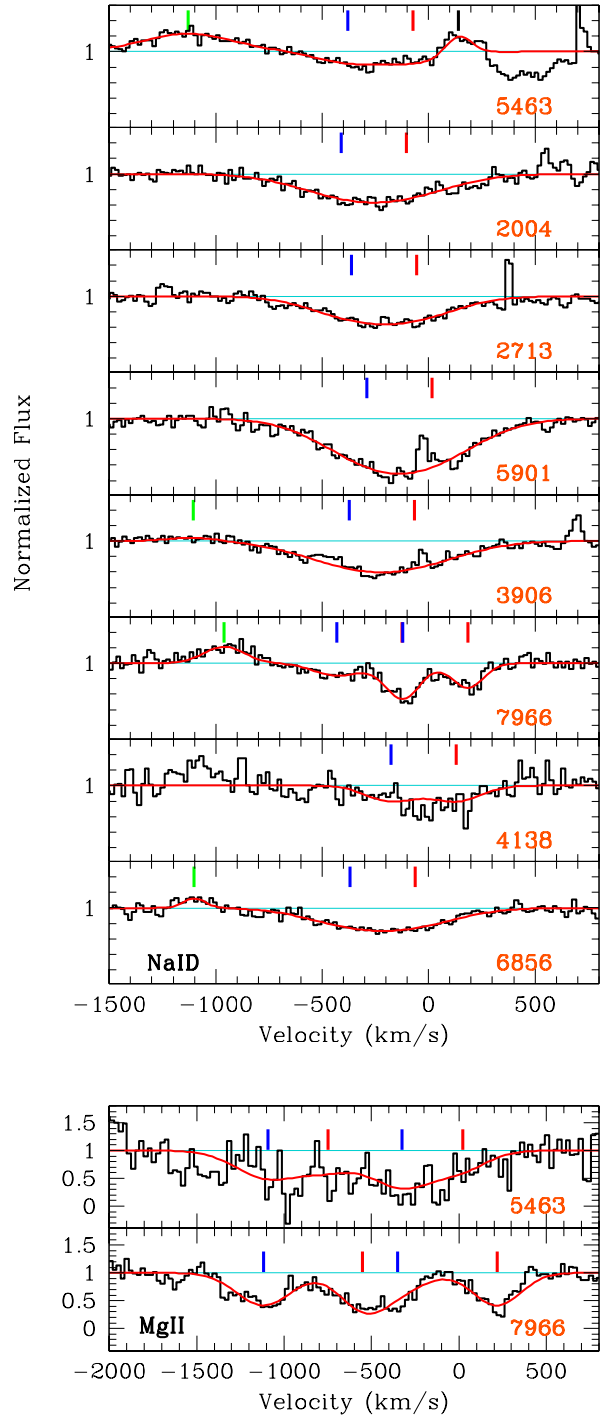}
\caption{{\NaID} (top) and {\MgII} (bottom) absorption intrinsic
  (down-the-barrel) to the AGN host galaxies. The flux-normalized
  spectra (black) are shown with respect to the AGN systemic velocity
  with $v=0$~{\kms}, set to the {\NaID} $\lambda$5876 line (top) and
  the {\MgII} $\lambda$2803 line (bottom). The wind model fitted to
  the data is shown along with red and blue tick-marks that indicate
  the red and blue components of the doublet required in the fit.  In
  four cases, a fit to the {\HeII} emission-line was required (green
  tick mark). The AGN name is listed in the bottom right corner of
  each panel. The results of the model fits are shown in
  Table~\ref{tab:windresults}. }
\label{fig:outflows}
\end{figure}

In Figure~\ref{fig:outflows}, we present the {\NaID} absorption
doublet for 8 of the 14 AGNs. Not all 14 absorption profiles are
presented since some are blended with strong sky features, such as the
$6860-6890$~{\AA} B-band, while others were not recorded because they
reside outside the spectral and spatial range of the spectrograph.
The flux normalized spectra are plotted at the systemic velocity of
the AGN, determined from emission-lines, relative to the {\NaID}
$\lambda$5897 line. We further show two intrinsic {\MgII} absorption
systems where the signal-to-noise ratio was sufficient to provide a
detection. Note that there is no significant evidence for {\MgII}
being detected in emission.  Qualitatively, in most cases, there
appears to be blueshifted outflows with maximum velocities of
$\sim$500~{\kms}.

To quantify the outflow velocities of the gas, we modeled our data
with a procedure similar to that of \citet{rupke05a}. Least-squares
Gaussian deblending was performed \citep[using the program
  FITTER;][]{archiveI} to estimate the equivalent widths, velocity
widths, and velocity centroids of component structures in the
absorption profiles. We fit the minimum number of Gaussian velocity
components to the absorption profiles that was
well-constrained/required by the data; typically one-to-two components
was sufficient to model each absorption system. For four of the
{\NaID} systems, we were required to simultaneously model the {\HeI}
$\lambda$5876 emission-line using a single component set to the
systemic velocity of the galaxy. In all cases we fixed the {\NaID} and
{\MgII} doublet ratios to unity to avoid unphysical models.

Our fit results are presented in Table~\ref{tab:windresults} and
Figure~\ref{fig:outflows}. The red and blue tick-marks in the figure
indicate the red and blue components of the doublet required in the
fit. Only for three {\NaID} systems (and both {\MgII} systems) were
two doublets required by the data.

Following \citet{rupke05b}, we adopt the definition that $\Delta v >
50$~{\kms} is outflowing wind material.  From
Table~\ref{tab:windresults}, we find that 6/8 {\NaID} systems contain
blueshifted absorption with $\Delta v > 50$~{\kms}, indicating
outflows with velocities ranging between 55 to 125~{\kms}. In 2/8
systems, one has redshifted absorption and the other has absorption
near the systemic velocity. For all 6 outflow objects, we find a mean
Doppler parameter of $b=220\pm70$~{\kms}, which is consistent with the
range of values derived by \citet{rupke05c} for 20 Seyfert~II galaxies
at $z=0.148^{+0.14}_{-0.07}$ having $b=$232$^{+244}_{-119}$~{\kms}.

\citet{rupke05c} also define the maximum outflow velocity to be the
central velocity of the most blueshifted component minus half its
FWHM: $\Delta v_{max}=\Delta v - $FWHM$/2$.  The $\Delta v_{max}$ for
our sample is listed in Table~\ref{tab:windresults} and we find a mean
value for the 6 wind systems of $\Delta v_{max}=277\pm87$~{\kms},
which is also consistent with, but smaller than, the values determined
by \citet{rupke05c} of $\Delta v_{max}=456^{+330}_{-191}$~{\kms}.

We have applied generalized Kendall and Spearman rank correlation
tests between the outflow properties with the galaxy inclination
angle. We only find suggestive trends with inclination and the outflow
equivalent width (2.1$\sigma$) and inclination and $b$
(1.7$\sigma$). Additional galaxies are required to test the
significance of these correlations.

The two {\MgII} absorption systems also exhibit high velocity
outflowing gas, higher than what is detected for {\NaID} in the same
systems. The blueshifted velocity offset for the bluest {\MgII}
components for both systems is $\sim 400$~{\kms} from the AGN systemic
velocity, a factor of more than $\sim 4$ higher than the {\NaID}
outflows. The Doppler parameters are consistent with those found for
the {\NaID}. However, $\Delta v_{max}$ is a factor of two higher than
those of the {\NaID} systems.  This suggests that {\MgII} traces
warmer gas and is more sensitive to outflowing material than the
{\NaID} absorption.

\begin{table}
\begin{center}
  \caption{The foreground AGN down-the-barrel absorption
    properties. The table columns are (1) the AGN--quasar pair name,
    (2) the ion measured, (3) the rest-frame equivalent width, (4) the
    cloud velocity offset from AGN systemic velocity, (5) the Doppler
    parameter or velocity width of the cloud, (6) the maximum outflow
    velocity defined as $\Delta v_{max}=\Delta v - $FWHM$/2$.}
  \vspace{-0.5em}
\label{tab:windresults}
{\footnotesize\begin{tabular}{lccrrr}\hline
Pair & Ion &$W_r$ & $\Delta v$\phantom{000}&  $b$\phantom{0000} & $\Delta v_{max}$\\
     &     & (\AA)     & (\kms)    &  (\kms) &  (\kms) \\\hline
 6856  &  {\NaID} & 2.05$\pm$0.13  &  $-$62$\pm$16             & 311$\pm$30 &  $-$218$\pm$22   \\
 4138  &  {\NaID} & 1.48$\pm$0.13  &    130$\pm$17             & 209$\pm$28 &    26$\pm$22     \\
 7966  &  {\NaID} & 1.14$\pm$0.04  &    185$\pm$3\phantom{0}   & 103$\pm$4\phantom{0}  &   $\cdots$\phantom{000}         \\
       &  {\NaID} & 0.81$\pm$0.05  & $-$125$\pm$10             & 151$\pm$13 &  $-$201$\pm$12   \\
       &  {\MgII} & 3.73$\pm$0.10  &    161$\pm$10             & 194$\pm$8\phantom{0}  &   $\cdots$\phantom{000}         \\
       &  {\MgII} & 2.60$\pm$0.11  & $-$406$\pm$8\phantom{0}   & 222$\pm$10 &  $-$547$\pm$62    \\
 3906  &  {\NaID} & 3.02$\pm$0.15  &  $-$68$\pm$16             & 358$\pm$28 &  $-$247$\pm$21    \\
 5901  &  {\NaID} & 5.22$\pm$0.17  &     16$\pm$9\phantom{0}   & 342$\pm$17 &  $-$155$\pm$12     \\
 2713  &  {\NaID} & 2.43$\pm$0.09  &  $-$55$\pm$17             & 294$\pm$33 &  $-$202$\pm$24     \\
2004   &  {\NaID} & 2.69$\pm$0.14  & $-$104$\pm$15             & 339$\pm$27 &  $-$274$\pm$20     \\
5463   &  {\NaID} & 1.02$\pm$0.17  &  $-$73$\pm$44             & 209$\pm$62 &  $-$178$\pm$54     \\
       &  {\MgII} & 2.11$\pm$0.30  &  $-$37$\pm$7\phantom{0}   & 212$\pm$11 &   $\cdots$\phantom{000}         \\
       &  {\MgII} & 6.31$\pm$1.01  & $-$382$\pm$55             & 330$\pm$58 &   $-$517$\pm$9\phantom{0}      \\\hline

\end{tabular}}
\end{center}
\end{table}
%%%%%%%%%%%%%%%%%%%%%%%%%%%%%%%%%%%%%%%%%%%%%%%%%%%%%%%%%%%%%%%%%%%%%%%%%%%%%%
%%%%%%%%%%%%%%%%%%%%%%%%%%%%%%%%%%%%%%%%%%%%%%%%%%%%%%%%%%%%%%%%%%%%%%%%%%%%%%

\section{Discussion}
\label{sec:dis}

It is clear that there are large {\HI}/metal reservoirs surrounding
quasars because their gas covering fractions are 60--80\% out to
200~kpc
\citep{hennawi06,bowen06,prochaska09,tytler09,farina13,prochaska13,farina14}.
\citet{farina14} show that the covering fraction of {\MgII} absorption
around quasars is more elevated than that found for field galaxies. We
find here that AGN have lower covering fractions,
0.09$^{+0.18}_{-0.08}$, than quasars, 0.47$^{+0.16}_{-0.15}$, and
possibly lower than galaxies, 0.25$^{+0.11}_{-0.09}$ between $100 \leq
D \leq 200$~kpc.

The main difference between galaxies, AGN, and quasars is the mass and
energy output of their central massive black holes.
\citet{chelouche08} modeled the distribution of cool gas around
quasars and concluded that conical AGN-driven winds could heat the gas
to temperatures of $\sim 10^5$K and could photo-ionize and
photo-evaporate cool {\MgII} gas clouds out to a few hundred
kiloparsecs.  This is consistent with the observed anisotropic
distribution of gas around quasars
\citep[e.g,][]{farina14}. Quasar--quasar pairs exhibit high transverse
covering fractions while intrinsic {\MgII} absorption in a quasar's
own spectrum is rarely seen
\citep[e.g.][]{hennawi07,prochaska09,farina13,prochaska13,farina14}.
\citet{farina14} report zero intrinsic {\MgII} absorption detections
in 26 quasar spectra.  This is consistent with our observations of
intrinsic {\MgII} absorption in the quasar, documented in Table~A1,
where we show that the background quasars only yield intrinsic {\MgII}
absorption in 1/5 objects ($W_r(2796) < 0.14$~{\AA},
3$\sigma$). However, for our foreground AGN sample, we detect 2/2
objects with intrinsic absorption along with 8/8 systems containing
intrinsic {\NaID} absorption (shown in Figure~\ref{fig:outflows}).
This result indicates that cool gas clouds intrinsic to the AGN are
not destroyed by the ionizing radiation originating central massive
black holes and suggests that the observed anisotropic distribution of
gas around quasars and AGN is consistent with the AGN unification
scheme \citep[e.g.][]{Urry95} since gas within AGN-driven winds is
heated to high temperatures, while the CGM remains unaffected outside
of the wind region.

Given that AGN outflows are roughly conical, with half opening angles
ranging between $\sim30-70$~degrees
\citep{hjelm96,veilleux01,muller-Sanchez11}, one might expect an
anisotropic distribution of cool gas surrounding the AGN: absorption
would be detected outside the jet region and there would be a lack of
absorption inside the jet region. We do not find a statistically
significant azimuthal angle dependence for the {\MgII} covering
fraction around the AGN hosts, though a possible hint of one may
exist.  A caveat to our geometric assumption, contrary to
starburst-driven winds, is that AGN outflow jets (Seyfert Is and IIs)
measured at radio wavelengths are oriented almost randomly relative to
the major axis of the host galaxy
\citep[e.g.][]{kinney00,gallimore06}.  Simulations have shown that
offsets between the jet orientation with respect to the angular
momentum of the disk can be caused by galaxy mergers or by
instabilities (gravitational or accretion) in isolated disks
\citep{hopkins12}. This could be why we do not find a strong trend
between the covering fraction and azimuthal angle.  Thus, if there is
a trend between covering fraction and orientation, it could be diluted
by the jet/disk offsets and thus detecting a statistically significant
trend (if it exists) would require a larger sample than the one we
present here.
 
It is well known that most optically-selected AGN also host nuclear
starbursts that produce starburst-driven winds
\citep[e.g.][]{veilleux01,rupke05c}.  Our host galaxies have a median
inclination of $i=51\pm12$ degrees and thus, if there are star
formation driven winds, we may observe them via blueshifted
absorption-line outflows (see Fig~\ref{fig:outflows}).  In 6/8 systems
we find {\NaID} outflows with $\Delta v > 50$~{\kms}, the maximum
blueshfited velocities reaching up to $\Delta
v_{max}=277\pm87$~{\kms}. These values are consistent with the
measurements made using {\NaID} absorption for 20 AGN Seyfert~IIs at a
similar redshift by \citet{rupke05c}. It is likely that these outflows
are primarily due to star-formation driven winds because the AGN winds
are perpendicular to the line-of-sight, although there could be a
combination of both \citep{veilleux01}.  \citet{rupke05c} find that
the star formation rates of Seyfert IIs range between
$118^{+151}_{-66}$M$_{\odot}$~yr$^{-1}$, with outflow masses and
outflow rates of log($M$/M$_{\odot}$)=8.8$\pm0.5$ and
$dM/dt=18^{+50}_{-13}$M$_{\odot}$~yr$^{-1}$, respectively.  Given the
consistencies between the outflow properties of \citet{rupke05c} and
our sample, we expect that our AGNs could have similar outflow masses
and rates produced in starburst-driven winds.

In two cases, shown in Figure~\ref{fig:outflows}, we also detect
{\MgII} absorption outflowing from the AGN at speeds $\sim$4 times
higher than found for {\NaID}. All of the detected {\NaID} and {\MgII}
absorption indicates the presence of gas reservoirs with most of them
outflowing from the host galaxy. However, even though we find that the
AGN hosts contain significant amounts of cool gas, we do not observe
it in their halos in {\MgII} absorption as probed by the background
quasars.

Given the high covering fraction found for quasars and the lack of
metal-line absorption within intrinsic down-the-barrel quasar spectra,
which would be equivalent to looking down the jet of an AGN, is highly
suggestive that the cool gas is ionized by the jets.  Furthermore, we
find evidence that cool gas is indeed being ejected at a few hundred
{\kms} from the AGN. However we do not observe significant amounts of
absorption around them. This adds credence to that idea that the
AGN-driven winds may be destroying the {\MgII}, consistent with the
AGN unified model \citep{Urry95}.

\section{Conclusions}
\label{sec:conclusion}

We have performed a detailed study of the cool gas covering fraction,
traced by {\MgII} absorption, surrounding narrow-line AGN
(Seyfert~IIs) using background quasar sight-lines. We analyzed 14
AGN--quasar pairs with the foreground AGN having a redshift range of
$0.12 \leq z \leq 0.22$.  The background quasars probe cool halo gas
surrounding the foreground AGNs over an impact parameter range of
$60\leq D \leq 265$~kpc. We further study the `down-the-barrel'
outflow properties of the AGNs themselves.

Our mains results can be summarized as follows: 

\begin{enumerate}
\item We find that AGN appear to have lower covering fractions between
  $100 \leq D \leq 200$~kpc, 0.09$^{+0.18}_{-0.08}$, than quasars,
  0.47$^{+0.16}_{-0.15}$, and possibly lower than galaxies,
  0.25$^{+0.11}_{-0.09}$. We do not find a statistically significant
  azimuthal angle dependence for the {\MgII} absorption covering
  fraction around the AGNs, though there could be a possible hint of
  one. More AGN--quasar pairs are required to demonstrate the
  azimuthal angle dependence or lack thereof and to test rigorously
  test for differences in the covering fractions.

\item AGN intrinsic {\NaID} absorption lines detected in 8/8 systems
  indicate that the AGN hosts have significant reservoirs of cool
  gas. This is validated by the 2/2 intrinsic {\MgII} absorption lines
  systems also detected.

\item We find that 6/8 intrinsic {\NaID} systems contain blueshifted
  absorption with $\Delta v > 50$~{\kms}, indicating outflows. For all
  6 outflow objects, we find a mean Doppler parameter of
  $b=220\pm70$~{\kms} and a maximum outflow velocity of $\Delta
  v_{max}=277\pm87$~{\kms}, both consistent with the range of values
  computed by \citet{rupke05c} for 20 Seyfert~II galaxies at a similar
  redshift.

\item The 2/2 intrinsic {\MgII} absorption systems also exhibit high
  velocity outflowing gas and are a factor of than $\sim 4$ higher
  than the {\NaID} outflows and the $\Delta v_{max}$ being a factor of
  two higher than those of the {\NaID} systems.  This is suggestive
  that {\MgII} traces warmer gas that is more sensitive to outflowing
  material than the {\NaID} absorption.

\end{enumerate}

Our results are consistent with AGN-driven winds ionizing the cool
gas, which dramatically decreases the cool $T \sim 10^4$K gas covering
fraction of AGN. Our observations also show that star-burst driven
winds are expelling cool gas into the CGM.  This picture is consistent
with results from quasar--quasar pair studies.  Previous
quasar--quasar studies have found that intrinsic cool gas
down-the-barrel is non-existent while their halos contain significant
amounts of cool gas. Both these results are complementary and provide
support for the AGN unified model.

%%%%%%%%%%%%%%%%%%%%%%%%%%%%%%%%%%%%%%%%
\section*{Acknowledgments}

We thank the anonymous referee for providing insightful comments and
improving the paper. MTM and JC thank the Australian Research Council
for Discovery Project grant DP130100568 and Future Fellowship grant
FT130101219 which supported this work. We thank W.M. Keck Observatory,
which is operated as a scientific partnership among the California
Institute of Technology, the University of California and the National
Aeronautics and Space Administration. Keck Observatory was made
possible by the generous financial support of the W.M. Keck
Foundation.  Data was also obtained from the Sloan Digital Sky Survey
(SDSS). Funding for the SDSS and SDSS-II has been provided by the
Alfred P. Sloan Foundation, the Participating Institutions, the
National Science Foundation, the U.S. Department of Energy, the
National Aeronautics and Space Administration, the Japanese
Monbukagakusho, the Max Planck Society, and the Higher Education
Funding Council for England. The SDSS Web Site is
http://www.sdss.org/.  The SDSS is managed by the Astrophysical
Research Consortium for the Participating Institutions. The
Participating Institutions are the American Museum of Natural History,
Astrophysical Institute Potsdam, University of Basel, University of
Cambridge, Case Western Reserve University, University of Chicago,
Drexel University, Fermilab, the Institute for Advanced Study, the
Japan Participation Group, Johns Hopkins University, the Joint
Institute for Nuclear Astrophysics, the Kavli Institute for Particle
Astrophysics and Cosmology, the Korean Scientist Group, the Chinese
Academy of Sciences (LAMOST), Los Alamos National Laboratory, the
Max-Planck-Institute for Astronomy (MPIA), the Max-Planck-Institute
for Astrophysics (MPA), New Mexico State University, Ohio State
University, University of Pittsburgh, University of Portsmouth,
Princeton University, the United States Naval Observatory, and the
University of Washington.

%{\it Facilities:} \facility{ARC (DIS)}, \facility{Keck II (ESI)},
%\facility{HST (WFPC--2)},\facility{VLT (UVES)}.

\appendix
\section{Additional Intrinsic and Foreground Absorbers}

We document additional intrinsic and foreground {\MgIIdblt} doublet
and {\CIVdblt} doublet absorption systems detected in our survey. For
intrinsic quasar absorption line systems we list the $3\sigma$
detection limit for an unresolved line located at the redshift of the
quasar.

\begin{table*}
\begin{center}
  \caption{Additional intrinsic and foreground {\MgIIdblt} doublet
    and {\CIVdblt} doublet absorption systems. The table columns are
    (1) the AGN name, (2) the AGN redshift, (3,4) the {\CIV} $\lambda
    1548$ and $\lambda 1548$ rest-frame equivalent width, (5) the
    {\CIV} $\lambda 1548$ absorption-line redshiftd, (6,7) the {\MgII}
    $\lambda 2796$ and $\lambda 2803$ rest-frame equivalent width, (8)
    and (10) the {\MgII} $\lambda 2796$ absorption-line redshift.}
  \vspace{-0.5em}
\label{tab:extra}
{\footnotesize\begin{tabular}{lccccccc}\hline
Name & $z_{QSO}$ & $W_r(1548)$ [\AA]& $W_r(1551)$ [\AA]& $z_{abs}$  & $W_r(2796)$ [\AA]&  $W_r(2803)$ [\AA]& $z_{abs}$ \\\hline  
9754 &  0.57072   &   $\cdots$       &      $\cdots$     &    $\cdots$  &     $\cdots$  &     $\cdots$    &     $\cdots$      \\
2004 &  1.05407   & $<$0.15          &    $<$0.15        &    $\cdots$  &    $<$0.05    &    $<$0.05      &     $\cdots$      \\
3273 &  1.57320   &   $\cdots$       &     $\cdots$      &    $\cdots$  &     $\cdots$  &     $\cdots$    &     $\cdots$      \\
3991 &  1.07677   & $<$0.08          &    $<$0.08        &    $\cdots$  &     $\cdots$  &     $\cdots$    &     $\cdots$      \\
5463 &  1.18228   & 0.26$\pm$0.02    &  0.19$\pm$0.02    &  1.1793724   & 0.24$\pm$0.01 &   0.13$\pm$0.01 &   1.1799584       \\
6856 &  0.52960   &   $\cdots$       &   $\cdots$        &    $\cdots$  &     $\cdots$  &     $\cdots$    &     $\cdots$      \\
     &            &  $\cdots$        &    $\cdots$       &    $\cdots$  & 0.58$\pm$0.04 &   0.56$\pm$0.04 &   0.2732928       \\
6863 &  0.24717   &  $\cdots$        &     $\cdots$      &    $\cdots$  &    $<$0.08    &    $<$0.08      &     $\cdots$      \\
1909 &  1.05373   &  $\cdots$        &     $\cdots$      &    $\cdots$  &    $<$0.14    &    $<$0.14      &     $\cdots$      \\
2713 &  1.35435   & 0.90$\pm$0.03    &  0.58$\pm$0.03    &  1.3550805   &     $\cdots$  &     $\cdots$    &     $\cdots$      \\
     &            & 0.79$\pm$0.03    &  0.66$\pm$0.03    &  1.1954113   &     $\cdots$  &     $\cdots$    &     $\cdots$      \\
3906 &  0.87571   & $\cdots$         &      $\cdots$     &    $\cdots$  &     $\cdots$  &     $\cdots$    &     $\cdots$      \\
4138 &  0.37035   & $\cdots$         &      $\cdots$     &    $\cdots$  &    $<$0.08    &   $<$0.08       &     $\cdots$      \\
5901 &  0.87540   & $\cdots$         &      $\cdots$     &    $\cdots$  &     $\cdots$  &     $\cdots$    &     $\cdots$      \\
7839 &  1.30369   & 0.98$\pm$0.03    & 0.92$\pm$0.03     &  1.3140370   &     $\cdots$  &     $\cdots$    &     $\cdots$      \\
     &            & 0.48$\pm$0.03    & 0.28$\pm$0.03     &  1.3035057   &     $\cdots$  &     $\cdots$    &     $\cdots$      \\
7966 &  0.36189   &  $\cdots$        &    $\cdots$       &    $\cdots$  &     $\cdots$  &     $\cdots$    &     $\cdots$      \\\hline

\end{tabular}}
\end{center}
\end{table*}

\label{lastpage}
\end{document}